\documentstyle [12pt]{article}
\title{{\bf Multiscale Technicolor and $b\rightarrow s\gamma$ }}
\author{Gongru Lu, Yigang Cao, Zhaohua Xiong,\\
 Chongxing Yue and Zhenjun Xiao\\
\small Physics Department of Henan Normal University, Xinxiang, Henan, 
453002, P.R. China  }
\date{}

\begin{document}
\maketitle
\begin{abstract}	
Correction to the $b\rightarrow s\gamma$ branching ratio in the multiscale 
walking technicolor model (MWTCM) is examined. For the original MWTCM, the 
correction is too large to explain the recent CLEO data. We show that if 
topcolor is further introduced, the branching ratio in the topcolor assisted 
MWTCM can be in agreement with the CLEO data for a certain range of the 
parameters.
\end{abstract}

\vspace{0.5cm}


\vspace{0.5cm}

\newpage
\begin{flushleft}    
\section*{I. Introduction}
\end{flushleft}
 Recently the CLEO collaboration has observed [1] the exclusive radiative 
decay $B\rightarrow K^*\gamma$ with a branching fraction of $BR ( B
\rightarrow K^*\gamma ) = ( 4.5 \pm 1.0 \pm 0.9 ) \times 10^{-5}$. The 
inclusive $b\rightarrow s\gamma$ branching ratio measured by CLEO [2] is
\begin{equation}
BR ( B\rightarrow X_s\gamma ) = ( 2.32 \pm 0.57 \pm 0.35 ) \times 10^{-4}.
\end{equation}
The newest upper and lower limits of this decay branching ratio are
\begin{equation}
1.0 \times 10^{-4} < BR ( B\rightarrow X_s\gamma ) < 4.2 \times 10^{-4}, at 
\ \ 95 \%  C.L.
\end{equation}
 As a loop-induced flavor changing neutral current ( FCNC ) process the 
inclusive decay ( at quark level ) $b\rightarrow s\gamma$ is in particular 
sensitive to contributions from those new physics beyond the standard model 
( SM ) [3]. There is a vast interest in this decay.

The decay $b\rightarrow s\gamma$ and its large leading log QCD corrections 
have been evaluated in the SM by several groups [4]. The reliability of the 
calculations of this decay is improving as partial calculations of the next-
to-leading logarithmic QCD corrections to the effective Hamiltonian [5,6]

The great progress in theoretical studies and in experiments achieved 
recently encourage us to do more investigations about this decay in 
technicolor theories.

Technicolor ( TC ) [7] is one of the important candidates for the mechanism 
of naturally breaking the electroweak symmetry. To generate ordinary fermion 
mass, extended technicolor ( ETC ) [8] models have been proposed. The 
original ETC models suffer from the problem of predicting too large flavor 
changing neutral currents ( FCNC ). It has been shown, however, that this 
problem can be solved in walking technicolor ( WTC ) theories [9]. 
Furthermore, the electroweak parameter $S$ in WTC models is smaller than that 
in the simple QCD-like ETC models and its deviation from the standard model 
( SM ) value may fall within current experimental bounds [10]. To explain the 
large hierarchy of the quark masses, multiscale WTC models ( MWTCM ) are 
further proposed [11]. These models also predict a large number of 
interesting  Pseudo-Goldstone bosons ( PGBs ) which are shown to be 
testable in future experiments [12]. So it is interesting to study physical 
consequences of these models.

In this paper, we examine the correction to the $b\rightarrow s\gamma$ decay 
from charged PGBs in the MWTCM. We shall see that the original MWTCM gives 
too large correction to the branching ratio of $b\rightarrow s\gamma$ due to 
the smallness of the decay constant $F_Q$ in this model. We shall show that 
if topcolor is further introduced, the branching ratio of $b\rightarrow s
\gamma$ in the topcolor assisted MWTCM can be in agreement with the CLEO data 
for a certain range of the parameters.

This paper is organized as the following: In Section II, we give a brief 
review of the MWTCM and then calculate the PGBs corrections to $b\rightarrow 
s \gamma$ decay, together with the full leading log QCD corrections. In 
Section III, we obtain the branching ratio of this decay. The conclusions 
and discussions are also included in this Section.    

\begin{flushleft}
\section*{II. Charged PGBs of MWTCM and QCD corrections to 
$b\rightarrow s\gamma$}
\end{flushleft}
 
Let us start from the consideration of the MWTCM proposed by Lane and Ramana 
[11]. The ETC gauge group in this model is
\begin{equation}
G_{ETC} = SU ( N_{ETC} )_1 \times SU ( N_{ETC} )_2,
\end{equation}
where $N_{ETC} = N_{TC} + N_C + N_L$ in which $N_{TC}$, $N_C$ and $N_L$ stand 
for the number of technicolors, the number of ordinary colors and the 
doublets of color-singlet technileptons, respectively. In Ref.[11], $N_{TC}$ 
and $N_L$ are chosen to be the minimal ones guaranteeing the walking of the 
TC coupling constant which are: $N_{TC} = N_L = 6$. The group $G_{ETC}$ is 
supposed to break down to a diagonal ETC gauge group $SU ( N_{ETC})_{1+2}$ at 
a certain energy scale. The decay constant $F_Q$ satisfies the following 
constraint [11]:
\begin{equation}
F= \sqrt{F^2_{\psi}+3F^2_Q+N_LF^2_L} = 246 GeV.
\end{equation}
It is found in Ref.[11] that $F_Q = F_L = 20 - 40 GeV$. We shall take 
$F_Q=40 GeV$ in our calculation. This present model predict the existence of 
a large number of PGBs, whose masses are typically expected to be larger than 
100 GeV. In this paper, we shall take the mass of color-singlet PGBs 
$m_{p^{\pm}}$ =100 GeV and the mass of color-octet PGBs $m_{p_8^{\pm}}$ = 
( 300 $\sim$ 600 ) GeV as the input parameters for our calculation. 

The phenomenology of those color-singlet charged PGBs in the MWTCM is similar 
with that of the elementary charged Higgs bosons $H^{\pm}$ of Type-I Two-
Higgs-Doublet model ( 2HDM ) [13]. And consequently, the contributions to the 
decay $b\rightarrow s\gamma$ from the color-singlet charged PGBs in the MWTCM 
will be similar with that from charged Higgs bosons in the 2HDM. As for the 
color-octet charged PGBs, the situation is more complicated because of the 
involvement of the color interactions. Other neutral PGBs do not 
contribute to the rare decay $b\rightarrow s\gamma$.     

The gauge coupling of the PGBs are determined by their quantum numbers. The 
Yukawa couplings of PGBs to ordinary fermions are induced by ETC interactions 
The relevant couplings needed in our calculation are
\begin{equation}
[ p^+--u_i--d_j ] = i\frac{1}{\sqrt{6}F_Q}V_{u_id_j}[m_{u_i}(1-\gamma_5)-
m_{d_j}(1+\gamma_5)],
\end{equation}
\begin{equation}
[ p_8^+--u_i--d_j ] = i\frac{1}{F_Q}V_{u_id_j}\lambda^a[m_{u_i}(1-\gamma_5)-
m_{d_j}(1+\gamma_5)],
\end{equation}
where $u = ( u, c, t )$, $d = ( d, s, b )$; $V_{u_id_j}$ is the corresponding 
element of Kobayashi-Maskawa matrix; $\lambda^a$ are the Gell-Mann 
$SU ( 3 )_c$ matrices.

 In Fig.1, we draw the relevant Feynman diagram which contributes to the 
decay $b\rightarrow s\gamma$, where the blob represents the photonic penguin 
operators including the $W$ 
gauge boson of the SM as well as the charged PGBs in the MWTCM. In the 
evaluation we at first integrate out the top quark and 
the weak $W$ bosons at $\mu=m_W$ scale, generating an effective five-quark 
theory. By using the renormalization group equation, we run the effective 
field theory down to b-quark scale to give the leading log QCD corrections, 
then at this scale, we calculate the rate of radiative b decay.

After applying the full QCD equations of motion [14], a complete set of 
dimension-6 operators relevant for $b\rightarrow s\gamma$ decay can be chosen 
to be
\begin{equation}
\begin{array}{l}
O_1 = ( \overline c_{L\beta}\gamma^{\mu}b_{L\alpha} )( \overline s_{L\alpha}
\gamma_{\mu}c_{L\beta} ),\\
O_2 = ( \overline c_{L\alpha}\gamma^{\mu}b_{L\alpha} ) ( \overline s_{L\beta}
\gamma_{\mu}c_{L\beta} ),\\
O_3 = ( \overline s_{L\alpha}\gamma^{\mu} b_{L\alpha} )\sum\limits_{
q=u,d,s,c,b} ( \overline q_{L\beta}\gamma_{\mu}q_{L\beta} ),\\
O_4 = ( \overline s_{L\alpha}\gamma^{\mu}b_{L\beta} )\sum\limits_{
q=u,d,s,c,b} (\overline q_{L\beta}\gamma_{\mu}q_{L\alpha} ),\\
O_5 = ( \overline s_{L\alpha}\gamma^{\mu}b_{L\alpha} )\sum\limits_{
q=u,d,s,c,b} ( \overline q_{R\beta}\gamma_{\mu}q_{R\beta} ),\\
O_6 = ( \overline s_{L\alpha}\gamma^{\mu}b_{L\beta} )\sum\limits_{
q=u,d,s,c,b}^{} ( \overline q_{R\beta}\gamma_{\mu}q_{R\alpha} ),\\
O_7 = ( e/16\pi^2 )m_b\overline s_L\sigma^{\mu\nu}b_{R}F_{\mu\nu},\\
O_8 = ( g/16\pi^2 )m_b\overline s_L\sigma^{\mu\nu}T^ab_RG^a_{\mu\nu}.
\end{array}
\end{equation}
The effective Hamiltonian at the $W$ scale is given as
\begin{equation}
H_{eff} = \frac{4G_F}{\sqrt{2}}V_{tb}V_{ts}^*\sum_{i=1}^{8}C_i(m_W)O_i(m_W).
\end{equation}
The coefficient of 8 operators are calculated and given as
\begin{equation}
\begin{array}{l}
C_i(m_W) = 0, i = 1, 3, 4, 5, 6, C_2 ( m_W ) = -1,\\ 
C_7(m_W)=\frac{1}{2}A(x)+\frac{1}{3\sqrt{2}G_FF_{Q}^2}[B(y)+8B(z)],\\     
C_8(m_W)=\frac{1}{2}C(x)+\frac{1}{3\sqrt{2}G_FF_{Q}^2}[D(y)
+(8D(z)+E(z))],
\end{array}
\end{equation}
with $x=(\frac{m_t}{m_W})^2$, $y=(\frac{m_t}{m_{p^{\pm}}})^2$, $z=
(\frac{m_t}{m_{p_8^{\pm}}})^2$. The functions $A$ and $C$ arise from graphs 
with $W$ boson exchange are already known contributions from the SM; while 
the functions $B$, $D$ and $E$ arise from diagrams with color-singlet and 
color-octet charged PGBs of MWTCM. They are given as
\begin{equation}
\begin{array}{l}
A(x)=-\frac{x}{12(1-x)^4}[(1-x)(8x^2+5x-7)+6x(3x-2)\ln x ],\\            
B(x)=\frac{x}{72(1-x)^4}[(1-x)(22x^2-53x+25)+6(3x^2-8x+4)\ln x ],\\                  
C(x)=-\frac{x}{4(1-x)^4}[(1-x)(x^2-5x-2)-6x\ln x ],\\
D(x)=\frac{x}{24(1-x)^4}[(1-x)(5x^2-19x+20)-6(x-2)\ln x ],\\
E(x)=-\frac{x}{8(1-x)^4}[(1-x)(12x^2-15x-5)+18x(x-2)\ln x ].
\end{array}
\end{equation}
The running of the coefficients of operators from $\mu=m_W$ to $\mu=m_b$ was 
well described in Refs.[4]. After renormalization group running we have the 
QCD corrected coefficients of operators at $\mu=m_b$ scale:  
\begin{equation}
C_7^{eff}(m_b) =\varrho^{-\frac{16}{23}}[C_7(m_W)+\frac{8}{3}
(\varrho^\frac{2}{23}-1)C_8(m_W)]+C_2(m_W)\sum\limits_{i=1}\limits^{8}h_i
\varrho^{-a_i},
\end{equation}
with 
$$
\begin{array}{l}
\varrho=\frac{\alpha_s(m_b)}{\alpha_s(m_W)},\\
h_i=(\frac{626126}{272277},-\frac{56281}{51730},-\frac{3}{7},-\frac{1}{14},
-0.6494,-0.0380,-0.0186,-0.0057),\\
a_i=(\frac{14}{23},\frac{16}{23},\frac{6}{23},-\frac{12}{23},0.4086,-0.4230,
-0.8994,0.1456).
\end{array}
$$

\begin{flushleft}
\section*{III.  The branching ratio of $B\rightarrow X_s\gamma$ and 
phenomenology}
\end{flushleft}

Following Refs.[4], applying the spectator model,
\begin{equation}
BR ( B\rightarrow X_s\gamma )/BR ( B\rightarrow X_ce\overline \nu )\approx 
\Gamma ( b\rightarrow s\gamma )/\Gamma ( b\rightarrow ce\overline \nu ). 
\end{equation}
If we take experimental result $BR ( B\rightarrow X_ce\overline\nu ) = 10.8
\%$ [15], the branching ratio of $B\rightarrow X_s\gamma$ is found to be
\begin{equation}
BR ( B\rightarrow X_s\gamma ) \approx 10.8\%\times\frac{\vert V_{tb}V_{ts}^*
\vert^2}{\vert V_{cb}\vert^2}\frac{6\alpha_{QED}\vert C_7^{eff}(m_b)\vert^2}
{\pi g(m_c/m_b)}(1-\frac{2\alpha_s(m_b)}{3\pi}f(m_c/m_b))^{-1}.
\end{equation}
Where the phase factor $g ( x )$ is given by
\begin{equation}
g ( x ) = 1 - 8x^2 + 8x^6 - x^8 - 24x^4 \ln x,
\end{equation}
and the factor $f (m_c/m_b )$ of one-loop QCD correction to the semileptonic 
decay is
\begin{equation}
f( m_c/m_b) = ( \pi^2 - 31/4 )( 1-m_c^2/m_b^2 ) + 3/2.  
\end{equation}

In numerical calculations we always use $m_W$ = 80.22 GeV, $\alpha_s(m_Z)$=
0.117, $m_c$ = 1.5 GeV, $m_b$ = 4.8 GeV and $\vert V_{tb}V_{ts}^*\vert ^2/
\vert V_{cb}\vert ^2$ = 0.95 [15] as input parameters.

Fig.2 is a plot of the branching ratio $BR ( B\rightarrow X_s\gamma )$ as a 
function of $m_{p_8^{\pm}}$ assuming $m_t$ = 174 GeV, $m_{p^{\pm}}$ = 100 
GeV.  The Long Dash line corresponds to the newest CLEO upper limit. From 
Fig.2 we can see that the contribution of the charged PGBs in the MWTCM is 
too large. In view of the above situation, we consider the topcolor assisted 
MWTCM. The motivation of introducing topcolor to MWTCM is the following:
In the MWTCM, it is very difficult to generate the top quark mass as large as 
that measured in the Fermilab CDF and D0 experiments [16], even with strong 
ETC [17]. Thus, topcolor interactions for the third generation quarks seem to 
be required at an energy scale of about 1 TeV [18]. In the present model, 
topcolor is taken to be an ordinary asymptotically free gauge theory, while 
technicolor is still a walking theory for avoiding large FCNC [19]. As in 
other topcolor assisted technicolor theories [19], the electroweak symmetry 
breaking is driven mainly by technicolor interactions which are strong near 1
TeV. The ETC interactions give contributions to all quark and lepton masses, 
while the large mass of the top quark is mainly generated by the topcolor 
interactions introduced to the third generation quarks. The ETC-generated 
part of the top quark mass is $m_t' = 66 k$, where $k \sim$ 1 to $10^{-1}$ 
[19]. In this paper, we take $m_t'$ = 15 GeV and 20 GeV as input parameters 
in our calculation ( i.e., in the above calculations, the $m_t$ =174 GeV is 
substituted for $m_t'$ = 15 GeV and 20 GeV, the other input parameters and 
calculations are the same as the original MWTCM ).  The $BR ( b\rightarrow 
s\gamma )$ in topcolor assisted MWTCM is illustrated in Fig.3. From Fig.3 
we can see that the obtained $BR ( b\rightarrow s\gamma )$ is in agreement 
with the CLEO data for a certain range of the parameters.

In this paper, we have not considered the effects of other possible 
uncertainties, such as that of $\alpha ( m_Z )$, next-to-leading-log QCD 
contribution [5], QCD correction from $m_t$ to $m_W$ [6] etc. The 
inclusion of those additional uncertainties will make the limitations weaken 
slightly.

\vspace{1cm}
\noindent {\bf ACKNOWLEDGMENT}

This work was supported in part by the National Natural Science 
Foundation of China, and by the funds from 
Henan Science and Technology  Committee.

\newpage
\begin {center}
{\bf Reference}
\end {center}
\begin{enumerate}
\item  
R. Ammar, $et \ \ al.$, CLEO Collaboration: Phys. Rev. Lett. 71 ( 1993 ) 674
\item
M. S. Alam, $et \ \ al.$, CLEO Collaboration: Phys. Rev. Lett. 74 ( 1995 ) 
2885
\item
J. L. Hewett, SLAC Preprint: SLAC-PUB-6521, 1994
\item
M. Misiak: Phys. Lett. B 269 ( 1991 ) 161; K. Adel, Y. P. Yao: Mod. Phys. 
Lett. A 8 ( 1993 ) 1679; Phys. Rev. D 49 ( 1994 ) 4945; M. Ciuchini $et \ \ 
al.$: Phys. Lett. B 316 ( 1993 ) 127
\item
A. J. Buras $et \ \ al.$: Nucl. Phys. B 370 ( 1992 ) 69; Addendum: ibid. B 
375 ( 1992 ) 501; B 400 ( 1993 ) 37 and B 400 ( 1993 ) 75; M. Ciuchini $et 
\ \ al.$: Phys. Lett. B 301 ( 1993 ) 263; Nucl. Phys. B 415 ( 1994 ) 403
\item
C. S. Gao, J. L. Hu, C. D. L \" u and Z. M. Qiu: Phys. Rev. D52 (1995)3978
\item
S. Weinberg: Phys. Rev. D 13 ( 1976 ) 974; D 19 ( 1979 ) 1277; L. Susskind: 
Phys. Rev. 20 ( 1979 ) 2619
\item
S. Dimopoulos and L. Susskind: Nucl. Phys. B 155 ( 1979 ) 237; E. Eichten and 
K. Lane: Phys. Lett. B 90 ( 1980 ) 125
\item
B. Holdom: Phys. Rev. D 24 ( 1981 ) 1441; Phys. Lett. B 150 ( 1985 ) 301; 
T. Appelquist, D. Karabali and L. C. R. Wijewardhana: Phys. Rev. Lett. 57 (
1986 ) 957
\item
T. Appelquist and G. Triantaphyllou: Phys. Lett. B 278 ( 1992 ) 345; R. Sundr
um and S. Hsu: Nucl. Phys. B 391 ( 1993 ) 127; T. Appelquist and J. Terning: 
Phys. Lett. B 315 ( 1993 ) 139.
\item
K. lane and E. Eichten: Phys. Lett. B 222 ( 1989 ) 274; K. Lane and M. V. 
Ramana: Phys. Rev. D 44 ( 1991 ) 2678
\item
E. Eichten and K. Lane: Phys. Lett. B 327 ( 1994 ) 129; V. Lubicz and P. 
Santorelli: BUHEP-95-16
\item
J. F. Gunion and H. E. Haber: Nucl. Phys. B 278, ( 1986 ) 449
\item
H. D. Politzer: Nucl. Phys. B 172 ( 1980 ) 349; H. Simma: Preprint, DESY 93-
083
\item
Particle Data Group: Phys. Rev. D 50 ( 1994 ) 1173
\item
F. Abe, $et \ \ al.$, The CDF Collaboration: Phys. Rev. Lett. 74 ( 1995 ) 
2626; S. Abachi, $et \ \ al.$, The D0 Collaboration: Phys. Rev. Lett. 74 ( 
1995 ) 2697
\item
T. Appelquist, M. B. Einhorn, T. Takeuchi, and L. C. R. Wijewardhana: Phys. 
Lett. B 220 ( 1989 ) 223; R. S. Chivukula, A. G. Cohen and K. Lane: Nucl. 
Phys. B 343 ( 1990 ) 554; A. Manohar and H. Georgi: Nucl. Phys. B 234 ( 
1984 ) 189
\item
C. T. Hill: Phys. Lett. B 266 ( 1991 ) 419; S. P. Martin: Phys. Rev. D 45 
( 1992 ) 4283; D 46 ( 1992 ) 2197; Nucl. Phys. B 398 ( 1993 ) 359; M. Linder 
and D. Ross: Nucl.Phys. B 370 ( 1992 ) 30; C. T. Hill, D. Kennedy, T. Onogi 
and H. L. Yu: Phys. Rev. D 47 ( 1993 ) 2940; W. A. Bardeen, C. T. Hill and 
M. Linder: Phys. Rev. D 41 ( 1990 ) 1649
\item
C. T. Hill: Phys. Lett. B 345, ( 1995 ) 483; K. Lane and E. Eichten, BUHEP-
95-11, hep-ph/9503433: To appear in Phys. Lett. B 
\end{enumerate}

\newpage

\begin{center}
{\bf Figure captions}                    
\end{center}

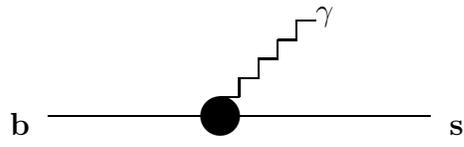
Fig.1: The Feynman diagram which contributes to the rare radiative decay $
b\rightarrow s\gamma$. The blob represents the photonic penguin operators 
including the $W$ gauge boson of the SM as well as the charged PGBs in 
the MWTCM.

Fig.2: The plot of the branching ratio of $ b\rightarrow s\gamma$ versus the 
mass of charged color-octet PGBs $m_{p_8^{\pm}}$ assuming $m_t$ = 174 GeV 
and the mass of color-singlet PGBs $m_{p^{\pm}}$ =
100 GeV in the MWTCM ( Solid line ). The Long Dash line corresponds to the 
newest CLEO upper limit. 

Fig.3: The plot of the branching ratio of $b\rightarrow s\gamma$ versus the   
mass of charged color-octet PGBs $m_{p_8^{\pm}}$ assuming the color-singlet 
PGBs $m_{p^{\pm}}$ = 100 GeV in the topcolor assisted MWTCM. The Solid line 
represents the plot assuming $m_t'$ = 15 GeV, and the Dot Dash line 
represents the plot assuming $m_t'$ = 20 GeV. The Long Dash line and 
Short Dash line correspond the newest CLEO upper and lower limits, 
respectively.
\newpage

\begin{picture}(30,0)
{\bf 
\setlength{\unitlength}{0.1in}
\put(15,-15){\line(1,0){20}}
 \put(13,-16){b}  
  \put(36,-16){s}
   \multiput(24,-15)(1,1){5}{\line(0,1){1}} 
   \multiput(23,-15)(1,1){6}{\line(1,0){1}}
    \put(24,-15){\circle*{3}}
\put(29,-10){$\gamma$} 
\put(24,-30){Fig.1} 
}
\
\end{picture}
\end{document}